\documentclass[12pt]{article}

\usepackage{newtxtext,newtxmath}

\usepackage{graphicx}

\usepackage[letterpaper,margin=1in]{geometry}

\renewenvironment{abstract}
	{\quotation}
	{\endquotation}

\date{}

\makeatletter
\renewcommand{\fnum@figure}{\textbf{Figure \thefigure}}
\renewcommand{\fnum@table}{\textbf{Table \thetable}}
\makeatother

\usepackage{scicite}

\usepackage{url}

\def\scititle{
	All-optical deconvolution via quantum cross-correlation spectroscopy with bright twin beams
}
\title{\bfseries \boldmath \scititle}

\author{
	Yujiro Eto$^{1\ast}$
	\and
	\small$^{1}$
	Center for Science Adventure and Collaborative Research Advancement (SACRA), 
	\\ \small Kyoto University, Kyoto
 	606-8502, Japan.\and
	\small$^\ast$Corresponding author. Email: eto.yujiro.8r@kyoto-u.ac.jp
}

\begin{document} 

\maketitle

\begin{abstract} \bfseries \boldmath
The resolution of spectroscopic measurements is fundamentally limited by the convolution between a system's intrinsic response and the instrument response function, creating an ill-posed inverse problem.
While entangled photons can theoretically bypass these limits, low photon flux has hindered their experimental realization.
Here, we propose quantum cross-correlation spectroscopy (QCCS), utilizing macroscopic twin beams (signal and idler beams) to physically circumvent this convolution limit in coherent anti-Stokes Raman scattering. 
Employing the signal beam as a broadband probe, QCCS achieves all-optical deconvolution through a correlation-enhanced sum-frequency generation channel between the convoluted anti-Stokes and idler beams.
By subtracting the uncorrelated background using a 500-femtosecond temporal window, we successfully retrieve narrow Raman bands (8~cm$^{-1}$ and 15~cm$^{-1}$) completely obscured by a 75~cm$^{-1}$ probe bandwidth without numerical deconvolution or prior knowledge. 
Our results establish macroscopic quantum correlations as a robust resource for high-fidelity spectroscopy, paving the way for ultrashort-pulse-free operando studies under natural, incoherent excitation conditions.
\end{abstract}

\noindent
\section*{INTRODUCTION}
In many spectroscopic measurements, the intrinsic response of a sample is observed as a convolution with the instrument response function.
Under realistic constraints such as finite temporal resolution, spectral bandwidth, and signal-to-noise ratio, this convolution process constitutes a fundamental limitation on the fidelity of the measured spectroscopic information.
In particular, spectral smoothing imposed by the instrument response function obscures fine spectral lineshapes, making it difficult to unambiguously identify the physical properties of the system under study.
Attempts to retrieve the intrinsic response through numerical post-processing generally lead to mathematically ill-posed inverse problems. 
Such deconvolution procedures are inherently unstable against noise and often require regularization schemes or prior assumptions, which may introduce model-dependent bias and artifacts \cite{goodman2015statistical, bertero2021introduction}.

These limitations are particularly pronounced in coherent spectroscopies such as coherent anti-Stokes Raman scattering (CARS), where broadband ultrashort pulses are frequently employed to capture a wide range of vibrational modes simultaneously or to track ultrafast dynamics \cite{cheng2016coherent}. 
However, the broad spectral envelope of such pulses acts as a dominant  instrument response function, directly limiting the achievable spectral resolution by smoothing narrow Raman features. 
To mitigate this trade-off, strategies such as hybrid femtosecond/picosecond schemes \cite{Dedic:14}, spectral focusing \cite{10.1063/1.1768312}, and noise autocorrelation spectroscopy \cite{xu2008noise} have been developed.
While effective, these approaches often increase experimental complexity, requiring active pulse shaping, complex optical scanning, or extensive ensemble averaging. 
Consequently, there remains a strong demand for a fundamentally different approach that can intrinsically compensate for the instrument response function within the measurement process itself, without relying on numerical regularization.

Quantum-entangled photon pairs have long been anticipated to relax conventional time–-frequency resolution constraints imposed by classical measurement frameworks through their time and frequency correlations \cite{RevModPhys.88.045008, 10.1063/5.0169768, PhysRevA.93.023807}. 
In recent years, quantum spectroscopy has achieved important milestones, including the observation of single-photon absorption and emission processes in natural photosynthetic complexes using entangled photon pairs \cite{li2023single}. 
However, efforts to extend these quantum advantages to advanced nonlinear spectroscopic techniques, such as two-dimensional optical spectroscopy \cite{10.1063/5.0056808} and CARS \cite{zhang2022entangled}, have proven significantly more challenging to realize experimentally.
This stagnation originates primarily from the intrinsically low photon flux of conventional quantum light sources and the stringent constraints associated with single-photon coincidence-based measurement schemes. 
Such approaches inevitably confine the photon flux to the single-photon regime, thereby fundamentally precluding the high excitation intensities required to drive nonlinear optical processes. 
As a result, the benefits of quantum spectroscopy have largely remained limited to the observation of extremely weak phenomena, leaving the high-flux regime, essential for practical chemical analysis and the investigation of ultrafast molecular dynamics, effectively unexplored.

To overcome these limitations, we propose the quantum cross-correlation spectroscopy (QCCS) scheme, which utilizes bright quantum-correlated twin beams and sum-frequency generation (SFG) readout.
In stark contrast to coincidence counting, this architecture permits scaling the photon flux into the macroscopic regime. 
By utilizing the enhancement of the frequency-correlated channel in the SFG process \cite{PhysRevLett.57.2516, dayan2003coherentcontrolbroadbandsqueezed, Kopylov_2019, Eto:24}, QCCS enables all-optical deconvolution of the probe spectrum.
By applying QCCS to CARS, we successfully recovered narrow Raman spectral features with full width at half maximum (FWHM) linewidths on the order of 8–15 cm$^{-1}$ from spectra instrumentally broadened to 75 cm$^{-1}$ by the probe bandwidth. 
While operation in the macroscopic flux regime inevitably introduces an uncorrelated SFG contribution, we demonstrate that the deconvolved spectrum can be isolated by subtracting measurements performed inside and outside a 500 fs temporal correlation window. 
Furthermore, we reveal that the resulting SFG output is imprinted with femtosecond-scale temporal cross-correlations between the signal and idler beams, as well as the self-correlation of the idler beam, when nanosecond-duration twin beams
are nonetheless employed.
This observation suggests the potential for achieving femtosecond time-resolved spectroscopy without the need for ultrafast laser sources, hinting at a paradigm shift in our utilization of incoherent light. 

\subsection*{All-optical deconvolution enabled by frequency correlation}

To elucidate the mechanism of all-optical deconvolution, we introduce a phenomenological theoretical model of the frequency-domain interaction.
In our scheme, the signal beam from the twin beams serves as the probe field, which we denote by $E_{\rm pr}(\omega)$.
When a sample with a spectral response $S(\omega_{v})$ is measured with the broadband probe, the intrinsic spectral features are inevitably broadened.
The field generated by the sample is described by the following convolution integral:
\begin{equation}
E_{\mathrm{sample}}(\omega)
\propto
\int S(\omega_{v})\, E_{\rm pr}(\omega - \omega_{v})\, d\omega_{v}.
\end{equation}
In conventional spectroscopy using classical broadband light, this convolution obscures the fine spectral structure of $S(\omega_{v})$, rendering the recovery of narrow intrinsic linewidths an ill-posed inverse problem.

In QCCS, this blurred field ($E_{\mathrm{sample}}$) is mixed with the correlated idler field, which serves as the gating field and is denoted by $E_{\rm ga}(\omega')$, via an SFG process.
The general SFG field is given by the following double integral:
\begin{equation}
E_{\mathrm{sf}}(\omega_{\mathrm{sf}})
\propto
\iint
S(\omega_{v})\,
E_{\rm pr}(\omega_{\mathrm{sf}} - \omega_{v} - \omega')\,
E_{\rm ga}(\omega')\,
d\omega_{v}\, d\omega'.
\end{equation}
This integral contains numerous frequency combinations.
QCCS exploits quantum correlations to selectively enhance a specific frequency-mixing channel.

When SFG is induced using twin beams generated by parametric down-conversion (PDC) pumped by a monochromatic laser at frequency $\Omega_{e}$,
the channel that satisfies $\omega_s + \omega_i = \Omega_{e}$ is enhanced due to the negative frequency correlations of the twin beams \cite{PhysRevLett.57.2516, dayan2003coherentcontrolbroadbandsqueezed, Kopylov_2019, Eto:24}.
Here, $\omega_s$ and $\omega_i$ are the frequencies of the signal and idler modes at the PDC source, respectively.
In the present context, the signal photon constitutes the probe field ($\omega_s \equiv \omega_{\rm pr}$), while the idler photon constitutes the gating field
($\omega_i \equiv \omega_{\rm ga}$).
Accordingly, the SFG amplitude is effectively constrained by the following $\delta$-function:
\begin{equation}
E_{\rm pr}(\omega_{\rm pr})\, E_{\rm ga}(\omega_{\rm ga})
\big|_{\text{pair}}
\propto
\delta(\omega_{\rm pr} + \omega_{\rm ga} - \Omega_{e}).
\end{equation}
It is important to note that such a $\delta$-function constraint does not arise for broadband classical fields.

Crucially, the integration over the gating-field frequency $\omega'$ leverages this strict frequency constraint and effectively removes the convolution
imposed by the broadband probe spectrum, leading to 
\begin{equation}
E_{\mathrm{sf}}^{\rm pair}(\omega_{\mathrm{sf}})
\propto
S(\omega_{\mathrm{sf}} - \Omega_{e}).
\end{equation}
In other words, the SFG component detected through the pairwise-correlated channel
is a direct spectral replica of the intrinsic, unconvolved sample response
$S(\omega_{v})$, shifted by $\Omega_{e}$.
Thus, by exploiting frequency correlations of the twin beams,
QCCS realizes an all-optical deconvolution of the probe bandwidth
and achieves spectral super-resolution.

However, the above derivation relies on the non-trivial assumption that quantum frequency correlations are preserved during the light--matter interaction.
The present model is intentionally phenomenological and is meant to elucidate the essential mechanism of correlation-enabled deconvolution.
A more rigorous theoretical treatment based on quantum-optical light--matter interactions is provided in the Supplementary Information.
This assumption is examined experimentally in the following sections.

\subsection*{Experimental procedure}
To demonstrate the principle of QCCS, 
we constructed an experimental setup that integrates a bright twin-beam source, a CARS spectroscopic interaction, and an SFG-based readout system, as schematically illustrated in Fig.~1A.
The corresponding energy-level diagram of the process is shown in Fig.~1B.

Quantum-correlated twin beams were generated via PDC excited by a 532-nm nanosecond laser.
A 2.5-cm-long periodically poled lithium niobate crystal was used to generate high-intensity signal and idler beams.
The PDC process produces broadband spectra determined by the phase-matching bandwidth.
Figure~1C shows the spectrum of the signal beam, which serves as the broadband probe field in the present experiment.
The correlation properties of the twin beams can be characterized using SFG \cite{Eto:24}.
A detailed characterization is presented in a subsequent section.
For the CARS measurements, the periodically poled lithium niobate phase-matching condition was tuned such that the center-frequency difference between the signal and idler beams matched the vibrational resonance of polystyrene ($\sim 1000~\mathrm{cm^{-1}}$).

To simplify the optical layout, we adopted a degenerate CARS configuration in which the twin beams served multiple roles.
The signal beam was used as both the pump and probe fields, while the idler beam served as the Stokes field.
A key aspect of the experimental design was to isolate the deconvolution effect of QCCS from other correlation-based enhancements in the CARS interaction, such as spectral focusing \cite{7pvj-3dqt}.
To this end, the optical delay $\tau_{1}$ of the Stokes beam relative to the pump and probe beams was set to be sufficiently longer than the correlation time for the twin beams.
Under this condition, the sample interacts with effectively uncorrelated broadband fields, resulting in the generation of a broad anti-Stokes spectrum convolved with the probe bandwidth.
The sample was a polystyrene plate, chosen to evaluate the recovery of its characteristic Raman bands near $1000~\mathrm{cm^{-1}}$.
For reference, the spontaneous Raman spectrum of polystyrene is shown in Fig.~1D; it exhibits FWHM linewidths of approximately $6~\mathrm{cm^{-1}}$ and $9~\mathrm{cm^{-1}}$.

The broadband anti-Stokes beam generated from the sample was subsequently directed to the SFG readout stage.
Here, the idler beam was reused as the gating field for the SFG process.
A 2-mm-thick type-I $\beta$-barium borate (BBO) crystal was employed for SFG.
The temporal overlap between the anti-Stokes signal and the gating idler field was controlled by a second delay line $\tau_{2}$.
The generated sum-frequency spectrum was detected using a spectrometer.

\subsection*{Twin-beam correlation characterization}

To verify that the generated twin beams exhibit the temporal and frequency correlations
required for QCCS, we characterized their correlation properties by placing a
0.1-mm-thick BBO crystal at the sample position and directly measuring the sum-frequency intensity and spectrum (Fig.~2A).

By scanning the relative delay $\tau_{1}$ between the signal and idler beams, we evaluated the temporal correlation through the delay dependence of the sum-frequency intensity and evaluated the frequency correlation through the spectral shape of the sum-frequency light.
Figure~2B shows the integrated sum-frequency intensity as a function of $\tau_{1}$.
It reveals a clear maximum centered at $\tau_{1}=0$ ps, which reflects the temporal correlation
between the twin beams.

When the signal and idler beams temporally overlapped ($\tau_{1}=0$), the sum-frequency spectrum exhibited a distinct narrow spectral feature, as shown in Fig.~2C.
The linewidth of this feature was approximately $0.1$~nm, limited by the spectral resolution of the spectrometer.
This feature originates from the enhancement of the correlated frequency-mixing channel that satisfies
\begin{equation}
\omega_s + \omega_i = \Omega_{e},
\end{equation}
and represents a spectral replica of the narrow-band PDC excitation.

In contrast, when the relative delay was increased to $\tau_{1}=1$~ps, well beyond the temporal correlation time, the narrow spectral component was no longer observed,
leaving only a broad and featureless sum-frequency background. 
This behavior demonstrates that the sum-frequency correlation of the twin beams is confined to a femtosecond-scale temporal window, despite their nanosecond-duration envelopes.

\subsection*{Transfer of twin-beam correlations to anti-Stokes field}

A fundamental prerequisite for QCCS is that the quantum correlations initially present between the signal and idler beams are not destroyed by the nonlinear Raman interaction, but can be transcribed into the anti-Stokes field generated through the CARS process.
In particular, QCCS requires that the anti-Stokes field effectively inherit the annihilation-operator character of the signal field, such that the frequency correlation can in principle be established between the anti-Stokes and idler fields.

As a first step toward examining this prerequisite, we measured the intensity of the anti-Stokes field while scanning the delay $\tau_{1}$ between the signal and idler beams.
As shown in Fig.~3A, the integrated anti-Stokes intensity exhibits a clear maximum centered at $\tau_{1}=0$, with a temporal FWHM of 310~fs.
This observation confirms that the twin-beam correlation is not limited to the direct SFG interaction shown in Fig.~2B, but also plays a critical role in the CARS generation process itself. 
This is consistent with our previously reported observation of correlation-induced enhancement in CARS \cite{7pvj-3dqt}.
However, this intensity measurement alone does not establish whether the correlation is actually transferred to the anti-Stokes field.

To directly investigate whether the transferred correlation manifests itself as an observable correlation between the anti-Stokes and idler fields,
we next fixed the delay at $\tau_{1}=1.3$~ps and $2.0$~ps, both well beyond the intrinsic correlation time for the twin beams, and measured the SFG intensity
while scanning the readout delay $\tau_{2}$ between the anti-Stokes field and the gating idler beam.
The resulting delay-dependent SFG intensity, shown in Figs.~4A and 4B, exhibits two classes of peak structures with distinct characteristics.
One peak shifts linearly with the delay $\tau_{1}$, while the other appears at a fixed value of $\tau_{2}$ independent of $\tau_{1}$.
The $\tau_{1}$-dependent shifting peak, exhibiting a temporal FWHM of 500 fs, reflects the cross-correlation between the anti-Stokes and idler fields established through the CARS process.
In contrast, the stationary peak can be attributed to the autocorrelation of the idler field, which stems from its intrinsic thermal photon statistics.

The coexistence of these two features demonstrates that the temporal correlation properties of the twin beams are effectively transferred to the anti-Stokes field through the CARS process. 
This preservation of temporal structure implies that the corresponding frequency correlations required for QCCS are also likely maintained, supporting the expectation that the correlation-enhanced channel is experimentally accessible at the SFG readout stage.

\subsection*{Spectral super-resolution via all-optical deconvolution}

To demonstrate spectral super-resolution enabled by QCCS, we first established the resolution limit imposed by the probe bandwidth using conventional CARS measurements.
The intrinsic Raman response of the sample is convolved with the phase-matched spectral components of the probe field during the CARS process.
When $\tau_{1}$ is chosen to be much longer than the correlation time of the twin beams (blue curve in Fig.~3B), the anti-Stokes spectrum is broadened to approximately $75~\mathrm{cm}^{-1}$.
As a result, the two closely spaced Raman bands of polystyrene, with narrow linewidths as shown in Fig.~1D, cannot be resolved.
Similar spectral obscuration was observed even at $\tau_{1}=0$ (green curve in Fig.~3B), confirming that the resolution is limited by the probe bandwidth regardless of the delay.

We next implemented QCCS readout by compensating for the delay $\tau_{1}$ introduced prior to the CARS interaction.
Specifically, the delay was fixed at $\tau_{1}=1.3$~ps, well beyond the temporal correlation time, and the delay $\tau_{2}$ was adjusted to bring the anti-Stokes field and the gating idler field back into their temporal correlation window.
Under this correlation condition ($\tau_{2}\approx -\tau_{1}$), the measured sum-frequency spectrum exhibits two distinct spectral peaks, as shown in Fig.~4C.
Notably, these spectra were retrieved with an acquisition time of only 2.5 minutes each.
In contrast, when the delay is detuned from the correlation window (e.g., $\tau_{2}=1.0$~ps), the sum-frequency spectrum reduces to a smooth, featureless background.

We subtracted the uncorrelated SFG background from the spectral component satisfying the condition ($\tau_{2}\approx -\tau_{1}$).
As shown in Fig.~4D, the resulting spectrum shows excellent agreement with the spontaneous Raman spectrum of the same sample, recovering narrow Raman bands with full widths at half maximum of 8~cm$^{-1}$ and 15~cm$^{-1}$. 
This direct correspondence demonstrates that QCCS physically removes the probe bandwidth convolution and retrieves the narrow Raman features through an entirely optical process, without relying on numerical deconvolution or regularization.
We attribute the slight remaining spectral deviation from the spontaneous Raman reference  (6~cm$^{-1}$ and 9~cm$^{-1}$) to the finite linewidth of the twin-beam pump ($\lesssim 0.1$~nm).
This finite linewidth leads to a degradation of frequency correlations, implying that the ideal delta function in Eq.~(3) effectively possesses a finite width.

Notably, the dispersive line shapes commonly observed in high-resolution CARS due to interference between resonant and non-resonant contributions are not prominent under the present conditions. 
One possible explanation for this phenomenon is interference between different field-matter quantum pathways \cite{RevModPhys.88.045008}.
As detailed in the Supporting Information, our theoretical derivation confirms the presence of this pathway interference within the QCCS framework.
Qualitatively, this pathway interference renders the heterodyne term phase-sensitive. 
This mechanism could lead to the effective suppression of the dispersive features.
While this qualitative picture is plausible, detailed numerical simulations are required for a more quantitative verification, which we leave for future work.

\section*{Discussion}

In this study, we demonstrated that QCCS optically eliminates probe-induced convolution artifacts, enabling the reconstruction of narrow-linewidth Raman responses from broadband data without numerical deconvolution.
By selectively enhancing a specific SFG channel via the signal--idler frequency correlation, QCCS realizes an all-optical deconvolution that automatically filters out the spectral blurring inherent to broadband probes.
This capability fundamentally distinguishes QCCS from conventional numerical deconvolution, offering a robust physical solution to the ill-posed inverse problem of recovering intrinsic spectra from band-limited measurements.

Crucially, our results reveal that the femtosecond-scale temporal correlations of the bright twin beams, which manifest as synchronized intensity fluctuations, are preserved even through the high-intensity nonlinear CARS process.
The successful extraction of this timing information via SFG readout confirms that the twin-beam correlation functions as an internal timing tag that operates independently of the external pulse duration. 
This concept marks a significant departure from conventional incoherent pump--probe spectroscopies, which typically rely on the autocorrelation of a single stochastic field to provide a temporal reference \cite{PhysRevA.30.2525, PhysRevA.29.2286}.
In contrast, we experimentally verified that the cross-correlation between two distinct correlated fields is directly mapped onto the SFG output. 
This demonstrates that the relative timing of the inherent fluctuations can be controlled and read out, validating the practical utility of the internal timing tag.

This architecture suggests that QCCS can evolve into a measurement framework possessing both ``time tags'' (temporal correlation) and ``frequency tags'' (frequency correlation).
Consequently, the trade-off between time and frequency resolution, traditionally dictated by the Fourier transform limit of the excitation pulse, could potentially be relaxed. 
With the implementon of this cross-correlation-based internal timing tag, femtosecond temporal referencing becomes feasible even under incoherent excitation conditions. 
This paves a realistic way toward ultrashort-pulse-free operando spectroscopy, in which natural photochemical dynamics could be resolved with femtosecond precision without imposing artificial coherence on the sample.

Furthermore, the ability to perform these measurements in the high-flux regime distinguishes QCCS from recent quantum spectroscopy milestones, which rely on single-photon coincidence counting \cite{li2023single}.
By utilizing the macroscopic SFG effect as a correlation gate, QCCS bridges the gap between quantum optics and nonlinear spectroscopy.
Although this study focused on demonstrating the readout and transfer of correlations, and not on time-resolved measurements of dynamics per se, the establishment of the internal timing tag provides a new degree of freedom absent in conventional paradigms.
We anticipate that this foundation will accelerate the development of time- and frequency-resolved metrology using high-intensity quantum-correlated light.

We showed that QCCS optically circumvents the convolution limit imposed by broadband probes, reconstructing narrow Raman features that are otherwise obscured by the probe bandwidth in conventional CARS measurements. 
Experimental confirmation that signal–idler correlations are transferred to the anti-Stokes field proves that quantum correlations survive high-intensity nonlinear interactions.
These results establish QCCS as a new measurement principle with the potential to bypass fundamental spectroscopic constraints. 
In particular, the use of femtosecond twin-beam correlations as an internal timing tag suggests the possibility of achieving simultaneous time–frequency resolution beyond the Fourier limit. 
QCCS is expected to pioneer new frontiers in spectroscopy, ranging from ultrashort-pulse-free time-resolved Raman measurements to the study of photochemical dynamics under incoherent excitation.

\clearpage
\begin{figure}[h]
\centering
\includegraphics[width=0.8\textwidth]{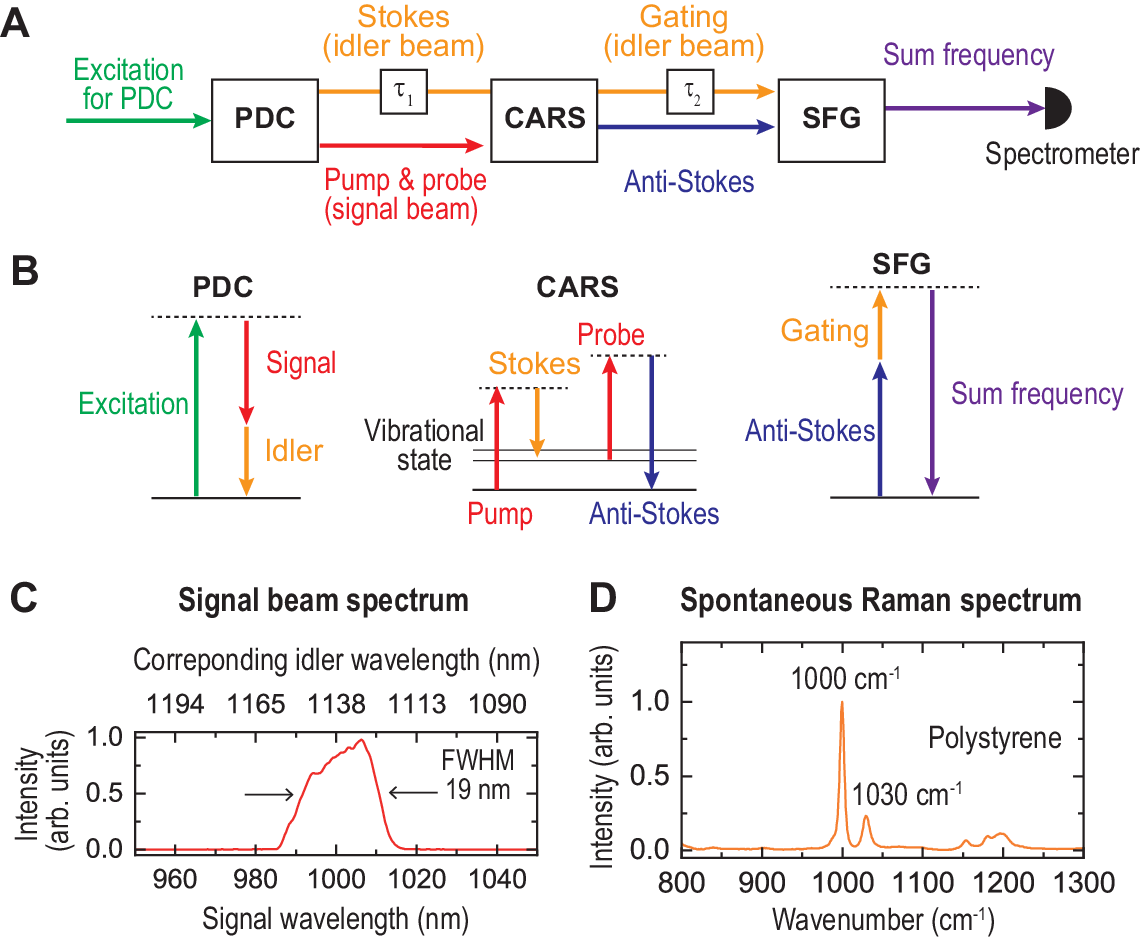}
\caption{
\textbf{Experimental framework for QCCS and spectral characteristics of the probe and Raman sample.}
(\textbf{A}) Schematic diagram of experimental setup that integrates a bright twin-beam source with a CARS system and an SFG readout stage.
(\textbf{B}) Corresponding energy-level diagrams for the PDC, CARS, and SFG processes.
(\textbf{C}) Spectrum of the signal beam used as the broadband probe field.
The FWHM of the signal spectrum is 19~nm.
The upper horizontal axis indicates the corresponding idler wavelength.
(\textbf{D}) Spontaneous Raman spectrum of polystyrene, showing characteristic Raman peaks at 1000~cm$^{-1}$ and 1030~cm$^{-1}$, which serve as reference vibrational modes for evaluating spectral resolution.
}
\label{fig1}
\end{figure}

\begin{figure}[h]
\centering
\includegraphics[width=0.4\textwidth]{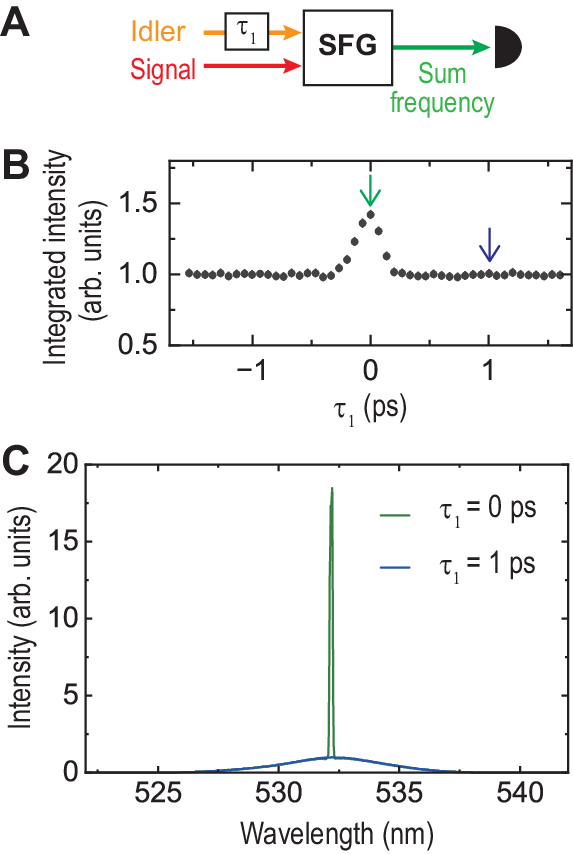}
\caption{
\textbf{Simultaneous verification of temporal and frequency correlations in bright twin beams.}
(\textbf{A}) Experimental configuration for correlation characterization, in which a BBO crystal (thickness 0.1~mm) is placed at the Raman sample position to directly measure sum-frequency generation between the signal and idler beams.
(\textbf{B}) Integrated sum-frequency intensity as a function of the relative delay $\tau_{1}$ between the signal and idler beams, showing a pronounced peak at temporal overlap ($\tau_{1}=0$).
The arrows indicate the delay positions corresponding to the spectra shown in panel (C).
(\textbf{C}) Sum-frequency spectra measured at $\tau_{1}=0$ and $\tau_{1}=1$~ps.
A sharp spectral peak is observed only at the temporal overlap, reflecting the enhancement of the frequency-correlated SFG channel.
}
\label{fig2}
\end{figure}

\begin{figure}[h]
\centering
\includegraphics[width=0.4\textwidth]{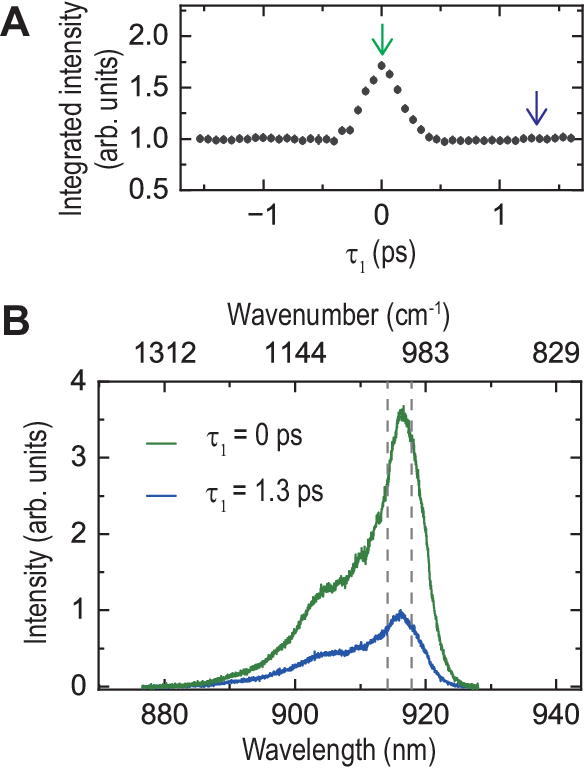}
\caption{
\textbf{Correlation-sensitive CARS process and spectral resolution limit imposed by probe bandwidth.}
(\textbf{A}) Integrated anti-Stokes intensity as a function of the delay $\tau_{1}$ between the signal and idler beams in the CARS interaction.
The intensity exhibits a clear maximum centered at $\tau_{1}=0$ with a temporal width of 310~fs.
(\textbf{B}) CARS spectra of a polystyrene sample measured with a broadband signal beam as the probe for delays $\tau_{1}=0$ and $\tau_{1}=1.3$~ps.
In both cases, the two intrinsic Raman bands are not spectrally resolved and appear as a single broadened feature due to convolution with the probe bandwidth.
The dotted lines indicate the positions of the two Raman modes. 
The lower horizontal axis shows the anti-Stokes wavelength and the upper axis shows the corresponding Raman shift calculated from the measured anti-Stokes wavelength as described in Methods.
}
\end{figure}

\begin{figure}[h]
\centering
\includegraphics[width=0.8\textwidth]{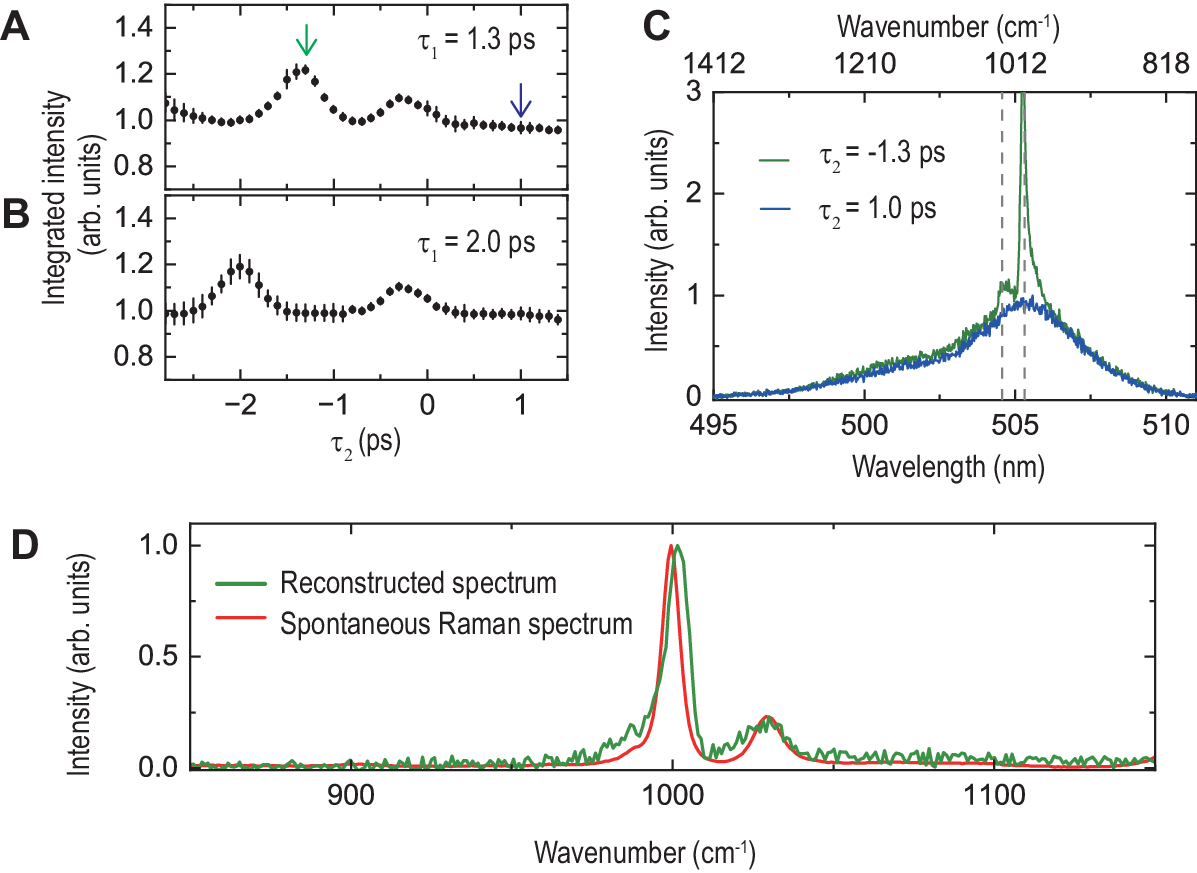}
\caption{
\textbf{Demonstration of all-optical spectral super-resolution.}
(\textbf{A, B}) Sum-frequency intensity as a function of the delay $\tau_{2}$ between the anti-Stokes field and the gating idler field, measured with the delay fixed at $\tau_{1}=1.3$~ps (A) and $\tau_{1}=2.0$~ps (B).
In each case, two distinct features are observed: a peak whose position shifts with $\tau_{1}$ and a peak that appears at a fixed value of $\tau_{2}$.
The shifting peak is attributed to  a correlation established between the anti-Stokes and idler fields through the CARS process,
whereas the stationary peak reflects the autocorrelation of the idler field.
(\textbf{C}) Sum-frequency spectra measured with $\tau_{1}=1.3$~ps under correlated ($\tau_{2} \approx -\tau_{1}$) and uncorrelated ($\tau_{2}=1.0$~ps) readout conditions.
When the delay $\tau_{2}$  compensates the initial delay $\tau_{1}$, two distinct spectral peaks emerge from the smooth broadband background.
(\textbf{D}) Reconstructed Raman spectrum (green curve) obtained by subtracting the uncorrelated SFG contribution from the spectrum, the latter of which was calculated by averaging the spectra at $\tau_2 = 1.3$~ps and $1.4$~ps. 
For comparison, the spontaneous Raman spectrum of the same sample is overlaid (red curve), demonstrating high-fidelity recovery of the narrow vibrational modes ($8$~cm$^{-1}$ and $15$~cm$^{-1}$) obscured by the $75$~cm$^{-1}$ probe bandwidth.
The lower horizontal axis shows the wavelength, and the upper axis indicates the corresponding Raman shift calculated as described in Methods.
}
\end{figure}


\clearpage 

%
\bibliography{science_template} 
\bibliographystyle{sciencemag}

%
%
%
%
%
%


\section*{Acknowledgments}
I would like to thank Prof. H. Kano for his valuable advice on the selection of Raman sample.
\paragraph*{Funding:}
This work was supported by the Japan Science and Technology Agency (JST) PRESTO (Grant No. JPMJPR2352), the Japan Society for the Promotion of Science (JSPS) KAKENHI (Grant No. 24K06919), and the RIKEN Transformative Research Innovation Platform (TRIP) initiative. 
\paragraph*{Author contributions:}
Y.E. performed the experiments, analyzed the data, and wrote the manuscript.
\paragraph*{Competing interests:}
Y.E. is an inventor on a patent application related to aspects of the spectroscopy technique described in this paper.
\paragraph*{Data and materials availability:}
All data needed to evaluate the conclusions in the paper are present in the paper


\subsection*{Supplementary materials}
Materials and Methods\\
Supplementary Text


\newpage


\renewcommand{\thefigure}{S\arabic{figure}}
\renewcommand{\thetable}{S\arabic{table}}
\renewcommand{\theequation}{S\arabic{equation}}
\renewcommand{\thepage}{S\arabic{page}}
\setcounter{figure}{0}
\setcounter{table}{0}
\setcounter{equation}{0}
\setcounter{page}{1} 


\begin{center}
\section*{Supplementary Materials for\\ \scititle}

Yujiro Eto$^{\ast}$
\small$^\ast$Corresponding author. Email: eto.yujiro.8r@kyoto-u.ac.jp
\end{center}

Materials and Methods\\
Supplementary Text
\newpage


\subsection*{Materials and Methods}

\subsubsection*{Twin-beam generation and CARS interaction}
The fundamental light source was a Q-switched laser that emitted pulses at a wavelength of 1064.4 nm with a FWHM of 3.9~ns and a repetition rate of
7.0~kHz (Cobolt, Tor 1064).
The fundamental beam was frequency-doubled via type-I second-harmonic generation in a 15-mm-long lithium triborate crystal (Edmund Optics, \#11-173).
The  lithium triborate crystal was housed in a custom-made oven and temperature-stabilized using a proportional--integral--derivative--controlled Peltier element (Thorlabs, TED200C).
The average output power of the second-harmonic beam at 532.2 nm was 150 mW. 
Residual fundamental light was removed using two short-pass filters (Thorlabs, FESH0750 and FESH0800).

The 532-nm beam was focused into a 2.5-cm-long periodically poled lithium niobate crystal (HC Photonics, SHVIS-SB-25) mounted in a temperature-controlled oven
(Covesion, PV40).
The beam waist diameter inside the crystal was set to 0.31~mm.
Bright quantum-correlated twin beams were generated via frequency-nondegenerate PDC in the high-gain regime.
The peak wavelengths of the signal and idler beams were 1007~nm and 1129~nm, respectively, with a total intensity of 2.8~mW.
The generated signal and idler beams were collimated using a lens with a focal length of 50~mm.
An iris was placed at the Fourier plane in a 2$f$ spatial-filtering configuration to suppress the spectral tails originating from non-collinear PDC components. 

To control the temporal delay $\tau_{1}$ between the twin beams, the signal and idler beams were spatially separated using a dichroic mirror (Semrock, Di02-R1064-25$\times$36).
The optical path length of the idler beam was adjusted using a retro-reflector mounted on a motorized linear translation stage (Thorlabs, MTS50/MZ8).
The two beams were subsequently recombined by the same dichroic mirror.
Similar implementations of the twin-beam source and timing control have been previously reported \cite{7pvj-3dqt, Eto:24, Eto:24ol}.

After recombination, the signal beam was spectrally filtered using a long-pass filter (Edmund Optics, \#86-231) to suppress weak short-wavelength spectral components that
make a negligible contribution to the CARS process but are detectable by the spectrometer.
The resulting signal spectrum after spectral filtering is shown in Fig.~1C. 
It was measured using a spectrometer (StellarNet Inc., Green-Wave NIR2b). 
The signal and idler beams were then focused by an objective lens (Mitutoyo, M Plan APO NIR 50$\times$) onto a 1.0-mm-thick polystyrene plate
(Kenis, PS 150-0664), and the generated anti-Stokes beam together with the co-propagating signal and idler beams was collimated using an objective lens
(Mitutoyo, M Plan APO NIR B 20$\times$).
The total intensity of the signal and idler beams immediately before the objective lens was 1.8~mW.

\subsubsection*{SFG readout}

After the CARS interaction, the generated anti-Stokes beam and the residual idler beam
were separated using a dichroic mirror (Thorlabs, DMLP1000).
Short-pass and long-pass filters were placed in the anti-Stokes and idler beam paths, respectively, to remove residual signal light.

To control the readout delay $\tau_{2}$ between the anti-Stokes field and the gating idler field, the optical path length of the gating beam was adjusted using a motorized translation stage in the same manner as for $\tau_{1}$.
The two beams were then recombined using a dichroic mirror (Thorlabs, DMSP1000) and focused by an achromatic lens with a focal length of 30~mm onto a 2-mm-long type-I BBO crystal (Castech Inc.) cut at $\theta = 22.9^\circ$ and $\phi = 0^\circ$.
The generated sum-frequency signal was spectrally resolved using a spectrometer (Andor Technology, Shamrock 500i) equipped with an EMCCD camera (Andor Technology, DU970P-BVF).
To optimize the signal-to-noise ratio, the EMCCD was operated at an electron-multiplying (EM) gain of 250.
For each spectrum, 15 accumulations were acquired with an exposure time of 10 s per frame, resulting in a total integration time of 150 s.

\subsubsection*{Definition of Raman shift for CARS spectra (Fig.~3B)}

The Raman shift $k$ (in cm$^{-1}$) shown on the upper axis in Fig.~3B was calculated from the measured anti-Stokes wavelength $\lambda_{\mathrm{as}}$. 
Since the signal and idler frequencies deviate from $\Omega_{e}/2$ by the same amount $\delta\Omega$ in opposite directions, the anti-Stokes frequency and the Raman shift follow the relations $\omega_{\mathrm{as}} \propto \Omega_{e}/2 + 3\delta\Omega$ and $k \propto 2\delta\Omega$, respectively.
 From these relationships, we derive the following conversion formula:
\begin{equation}
k = \frac{2}{3} \left[ \frac{1}{\lambda_{\mathrm{as}}} - \frac{1}{2\lambda_{\Omega_e}} \right],
\end{equation}
where $\lambda_{\Omega_e}$ corresponds to the wavelength of the excitation frequency $\Omega_e$ for the PDC source.

\subsubsection*{Definition of Raman shift for SFG readout spectra (Fig.~4C)}

The Raman shift $k$ (in cm$^{-1}$) shown on the upper axis in Fig.~4C was determined as follows. 
First, the SFG frequency $\omega_{\mathrm{sf}}$ is related to the vibrational frequency $2\delta\Omega$ by the linear relation $\omega_{\mathrm{sf}} = \Omega_{e} + 2\delta\Omega$. 
This implies that the Raman shift $k$ is directly mapped onto the frequency offset from the PDC excitation frequency $\Omega_e$. 
Based on this relationship, the Raman shift $k$ was calculated from the measured SFG wavelength $\lambda_{\mathrm{sf}}$ using the following conversion formula:
\begin{equation}
k = \frac{1}{\lambda_{\mathrm{sf}}} - \frac{1}{\lambda_{\Omega_e}}.
\end{equation}

\subsection*{Supplementary Text: quantum-optical description of quantum cross-correlation spectroscopy}

Here, we present a quantum-optical description of quantum cross-correlation spectroscopy (QCCS), explicitly accounting for the quantum properties of the broadband twin beams generated by parametric down-conversion (PDC), their interaction with matter through coherent anti-Stokes Raman scattering (CARS), and the subsequent sum-frequency generation (SFG) readout.

The purpose of this theoretical analysis is threefold:
(i) to provide a rigorous quantum description underlying the phenomenological frequency-domain model introduced in the main text;
(ii) to identify the correlation-enhanced SFG contribution responsible for the all-optical deconvolution of the probe bandwidth; and
(iii) to distinguish this contribution from uncorrelated SFG background terms that inevitably arise in the macroscopic, multimode regime.

Throughout this Supplementary Information, the optical fields are treated as operators, enabling a fully quantum description that remains valid in the macroscopic photon-number regime.
This approach allows us to explicitly track multi-operator correlation functions and to identify the specific frequency-mixing channels that give rise to the correlation-enhanced SFG underlying QCCS.

\subsubsection*{Quantum description of twin beams}

The notation used to describe the twin beams follows our previous work \cite{7pvj-3dqt}.
Here, we summarize only what is needed for the present discussion.

The annihilation operator describing the twin beams is given by
\begin{equation}
\hat{a}(\omega)
=
\mu(\omega)\hat{a}_{\rm v}(\omega)
+i\nu(\omega)\hat{a}_{\rm v}^{\dagger}(\Omega_{e}-\omega),
\label{eq:S1}
\end{equation}
where $\mu(\omega)$ and $\nu(\omega)$ are the parametric interaction coefficients, and $\Omega_{e}$ denotes the angular frequency of the monochromatic excitation field driving the PDC process.
For a type-0 PDC process, the symmetry relations $\mu(\omega)=\mu(\Omega_{e}-\omega)$ and $\nu(\omega)=\nu(\Omega_{e}-\omega)$ hold \cite{PhysRevA.87.013844}.

\subsubsection*{Quantum description of CARS followed by SFG}

We now present a quantum-optical formulation of CARS followed by SFG, where the latter constitutes the detection stage of QCCS.
Throughout this section, the subscripts $\mathrm{pu}$, $\mathrm{st}$, $\mathrm{pr}$, and $\mathrm{as}$ label the pump, Stokes, probe, and anti-Stokes fields involved in the CARS process, respectively.
The pump and probe fields correspond to the signal beam of the twin beams, while the Stokes field corresponds to the idler beam.
The subscript $\mathrm{ga}$ denotes the idler field employed to gate the subsequent SFG process.

\paragraph*{CARS-induced anti-Stokes field.}
We first derive the annihilation operator of the anti-Stokes field generated through the CARS interaction, 
which serves as the input to the subsequent SFG-based detection stage of QCCS.
The CARS-induced anti-Stokes field operator is given by
\begin{equation}
\begin{aligned}
\hat{a}_{\rm as}(\omega_{\rm as};\tau_{1})
\propto\;&
\int^{\Omega_{e}}_{\Omega_{e}/2} d\omega_{\rm pr}
\int^{\Omega_{e}/2}_{0} d\omega_{\rm st}
\int^{\Omega_{e}}_{\Omega_{e}/2} d\omega_{\rm pu}\,
\chi^{(3)}(\omega_{\rm pu}-\omega_{\rm st})\,
\hat{a}(\omega_{\rm pr})\,
\hat{a}^{\dagger}(\omega_{\rm st})\,
\\
&\times
e^{-i\omega_{\rm st}\tau_{1}}
\hat{a}(\omega_{\rm pu})\,
\delta(\omega_{\rm pu}-\omega_{\rm st}+\omega_{\rm pr}-\omega_{\rm as}) .
\end{aligned}
\label{eq:CARS_as}
\end{equation}
where the probe delay $\tau_{1}$ enters through the linear phase factor associated with the Stokes field.
In this notation, the spectral phases (i.e., phase dispersion) of all optical fields are neglected for simplicity.

\paragraph*{SFG field.}
We next define the SFG field, which constitutes the detection stage of QCCS.
In the frequency domain, the SFG field is generated by mixing the CARS-induced anti-Stokes field with the gating idler field.
It is expressed as
\begin{equation}
\hat{a}_{\rm sf}(\omega_{\rm sf};\tau_{1},\tau_{2})
\propto
\int^{\Omega_{e}/2}_{0} d\omega_{\rm ga}
\int d\omega_{\rm as}\,
\hat{a}(\omega_{\rm ga})\,e^{i\omega_{\rm ga}(\tau_{1}+\tau_{2})}\,
\hat{a}_{\rm as}(\omega_{\rm as};\tau_{1})\,
\delta(\omega_{\rm ga}+\omega_{\rm as}-\omega_{\rm sf}),
\label{eq:SFG_def}
\end{equation}
where $\tau_{2}$ denotes the additional delay applied to the gating idler field.
As the same idler field participates in both the CARS interaction and the subsequent SFG process,
the total delay accumulated by the idler field is given by $\tau_{1}+\tau_{2}$.

\paragraph*{Fourth-order coherence in SFG-based detection.}

The SFG spectrum is given by the normally ordered expectation value
\begin{equation}
S_{\rm sfg}(\omega_{\rm sf})
=
\langle 0 |
\hat{a}^{\dagger}_{\rm sf}(\omega_{\rm sf})
\hat{a}_{\rm sf}(\omega_{\rm sf})
| 0 \rangle ,
\end{equation}
which is expressed in terms of
fourth-order coherence functions of the twin beams.
This provides a quantum-optical description of the SFG detection process underlying the phenomenological frequency-domain model introduced in the main text.

In the experiment, the delay $\tau_{1}$ introduced in the CARS interaction
is chosen to be sufficiently large compared to the inverse bandwidth of the twin beams,
so as to effectively suppress signal--idler frequency correlations during the CARS process.
As a result, among the many possible fourth-order contraction terms,
only those containing the form
$\delta(\omega'_{\rm st}-\omega_{\rm st})$ survive the frequency integration,
where the primed frequency $\omega'$ denotes a frequency associated with the bra state in Eq.~(S6).

\subsubsection*{Correlation-enhanced and uncorrelated SFG channels}

Under the above condition, the fourth-order correlation functions accessed by the SFG process
can be classified into physically distinct contributions according to the frequency
constraints imposed by the surviving delta functions.

We first consider the correlation-enhanced SFG channel, which arises from
fourth-order coherences containing the frequency constraint
$\delta(2\Omega_{e}-\omega_{\rm pr}-\omega_{\rm ga})$
or its equivalent primed form.
One representative term of this class is, for example, given by
\begin{equation}
D^{(4)}_{\rm pair}
=
\delta(\Omega_{e}-\omega_{\rm pr}-\omega_{\rm ga})
\delta(\Omega_{e}-\omega'_{\rm pu}-\omega_{\rm ga})
\delta(\omega'_{\rm st}-\omega_{\rm st})
\delta(\omega'_{\rm pr}-\omega_{\rm pu}) .
\end{equation}
This frequency constraint implies the strict frequency anticorrelation between the probe and
the gating idler fields originating from PDC.
As a result, the frequency spread of the broadband probe field is compensated by that
of the gating idler, leading to an effective cancellation of the probe bandwidth.
This mechanism gives rise to the correlation-enhanced SFG and underlies
the all-optical deconvolution realized in QCCS.
Furthermore, the frequency constraints in $D^{(4)}_{\rm pair}$ lead the third-order susceptibility $\chi^{(3)}$ and its conjugate $\chi^{(3)*}$ to follow different frequency arguments. 
This mathematical structure represents interference between different field-matter quantum pathways.

In contrast, the uncorrelated SFG background terms do not fully retain the frequency correlations of the twin beams.
One representative term belonging to this uncorrelated class is given by 
\begin{equation}
D^{(4)}_{\rm bg}
=
\delta(\omega'_{\rm ga}-\omega_{\rm ga})
\delta(\omega'_{\rm pr}-\omega_{\rm pr})
\delta(\omega'_{\rm st}-\omega_{\rm st})
\delta(\omega'_{\rm pu}-\omega_{\rm pu}) .
\end{equation}
This term contains the factor $\delta(\omega'_{\rm ga}-\omega_{\rm ga})$.
As a result, by taking the total delay $\tau_{1}+\tau_{2}$ to be sufficiently large, it is possible to selectively identify this major background contribution.

An additional term involving correlation between the pump and the gating idler fields exists.
Although this term cannot be completely isolated, the spectrum arising from it is broadened by the broadband probe field and is therefore not expected to play a dominant role under the present experimental conditions.



\end{document}